\documentclass[10pt,twocolumn,letterpaper]{article}

\usepackage[pagenumbers]{cvpr} % To force page numbers, e.g. for an arXiv version

\usepackage{graphicx}
\usepackage{amsmath}
\usepackage{amssymb}
\usepackage{booktabs}
\usepackage{import}
\usepackage{float}
\usepackage{longtable}
\usepackage{multicol}
\usepackage[table]{xcolor}

\usepackage{amssymb}

\newcommand\pimencoded{PIM-encoded}
\newcommand\pimd{PIMD}
\newcommand\pimtool{PIMTool}
\newcommand\pimm{PIMM}

\usepackage[pagebackref,breaklinks,colorlinks]{hyperref}

\usepackage[capitalize]{cleveref}
\crefname{section}{Sec.}{Secs.}
\Crefname{section}{Section}{Sections}
\Crefname{table}{Table}{Tables}
\crefname{table}{Tab.}{Tabs.}

\newcommand\nnfootnote[1]{%
  \begin{NoHyper}
  \renewcommand\thefootnote{}\footnote{#1}%
  \addtocounter{footnote}{-1}%
  \end{NoHyper}
}

\def\cvprPaperID{6029} % *** Enter the CVPR Paper ID here
\def\confName{CVPR}
\def\confYear{2023}

\begin{document}

\title{PIM: Video Coding using Perceptual Importance Maps}

\author{Evgenya Pergament\footnotemark \\
Stanford University\\
\and
Pulkit Tandon\\
Stanford University\\
\and
Oren Rippel \\
WaveOne, Inc. \\
\and
Lubomir Bourdev \\
WaveOne, Inc. \\
\and
Alexander G. Anderson \\
WaveOne, Inc. \\
\and
Bruno Olshausen \\
University of California, Berkeley \\
\and
Tsachy Weissman \\
Stanford University \\
\and
Sachin Katti \\
Stanford University \\
\and
Kedar Tatwawadi\\
WaveOne, Inc. \\
}
\maketitle
\nnfootnote{*Corresponding author: evgenyap@stanford.edu}

%%%%%%%%% ABSTRACT
\begin{abstract}
Human perception is at the core of lossy video compression, with numerous approaches developed for perceptual quality assessment and improvement over the past two decades. In the determination of perceptual quality, different spatio-temporal regions of the video differ in their relative importance to the human viewer. However, since it is challenging to infer or even collect such fine-grained information, it is often not used during compression beyond low-level heuristics. We present a framework which facilitates research into fine-grained subjective importance in compressed videos, which we then utilize to improve the rate-distortion performance of an existing video codec (x264). The contributions of this work are threefold: (1) we introduce a web-tool which allows scalable collection of fine-grained perceptual importance, by having users interactively paint spatio-temporal maps over encoded videos; (2) we use this tool to collect a dataset with 178 videos with a total of 14443 frames of human annotated spatio-temporal importance maps over the videos; and (3) we use our curated dataset to train a lightweight machine learning model which can predict these spatio-temporal importance regions. We demonstrate via a subjective study that encoding the videos in our dataset while taking into account the importance maps leads to higher perceptual quality at the same bitrate, with the videos encoded with importance maps preferred $1.8 \times$ over the baseline videos. Similarly, we show that for the 18 videos in test set, the importance maps predicted by our model lead to higher perceptual quality videos, $2 \times$ preferred over the baseline at the same bitrate. 
\end{abstract}

\section{Introduction}
\label{sec:intro}
\begin{figure*}
\centering
\includegraphics[width=0.8\textwidth]{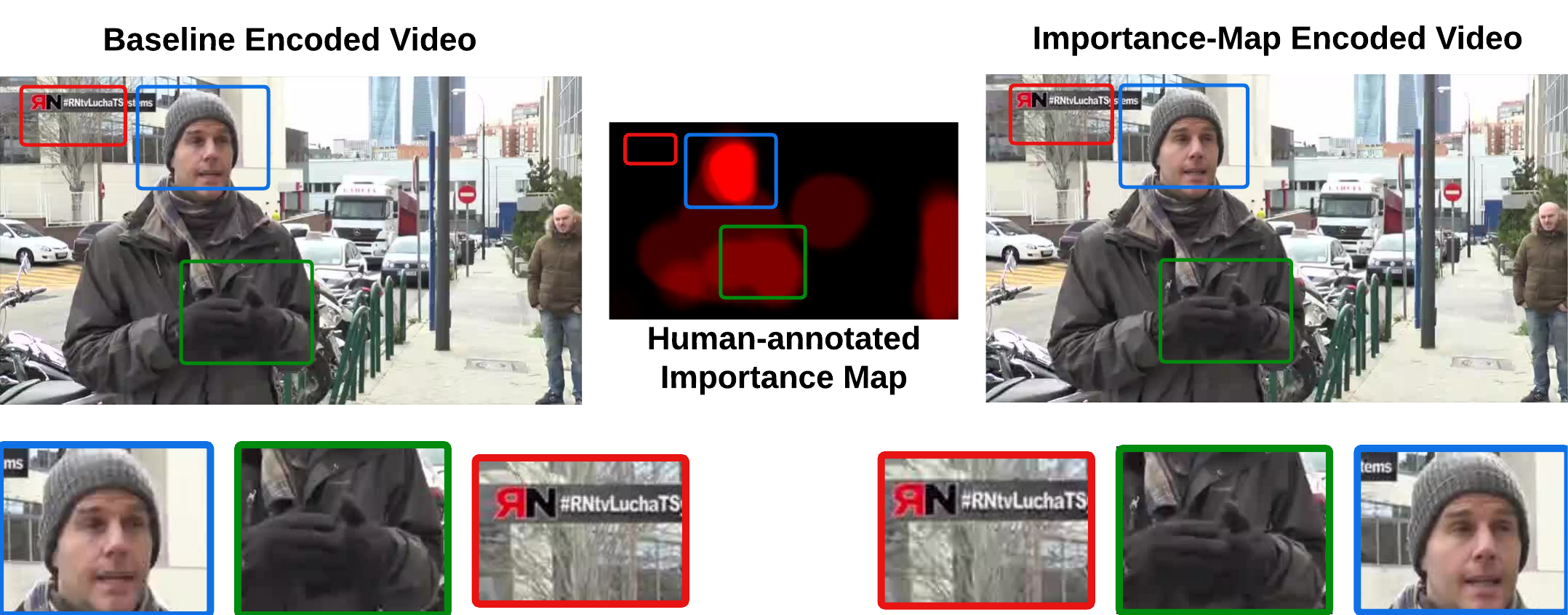}
\caption{An example of our perceptual importance map framework. {\bf Left}: a frame from a video in the Perceptual Importance Map Dataset (Sec. \ref{sec:dataset}), encoded using x264. {\bf Middle}: a human-annotated perceptual importance map collected using \pimtool{} (Sec. \ref{sec:annotation_tool}). {\bf Right}: the same frame, encoded at the same bitrate using a version of x264 modified to spatially modulate the bitrates/qualities using an input importance map. \pimtool{} enables collecting fine-grained importance maps, which allow improving perceptual quality by reallocating bits from less important to more important regions: in this example, the importance-map encoded video was preferred 6.25$\times$ over the baseline in a subjective study.  The blue and green bounding boxes show regions marked as important (at different levels), and the red bounding box shows the trade-off with another conventionally salient region which is less important in this case. See \cref{sec4:dataset:results:qualitative} for details.}
\label{sec4:fig:dataset_example}
\end{figure*}

The amount of video data generated, stored and shared has exploded exponentially over the past decade. Today, more than ever, video plays a key role in our lives with the plethora of streaming services and video conferencing solutions ubiquitous in our daily lives. Given the limited internet bandwidth, this has been made possible mainly by the consistent improvements to video compression technologies over the past few decades. 

Lossy video compression algorithms work by discarding less ``important'' information from the video and efficiently representing the retained information. A gargantuan challenge has been to quantitatively define importance to best correlate with the human visual system. This necessity has spurred decades of research on a variety of facets of the problem, ranging from how to conduct user studies for subjective quality evaluations, to designing metrics to assess perceptual quality, to utilizing the lessons learned towards perceptual compression (see review papers \cite{challenges_perceptual_compression, lee2012perceptual, zhang2021survey}).

In their contribution to perceptual quality, different pixels in a given video are not equal in significance: some are much more important than others, as distortions to the content they carry are more perceptible to the human viewers. 

How, then, do we assign importance values to each pixel in a video? One possibility is to produce scores using objective metrics such as PSNR or SSIM; however, these only capture low-level information without an actual understanding of the content. An alternative approach would be to compute saliency or object segmentation maps, which can be used as proxies to perceptual importance; these, however, fail to capture low-level issues (e.g. banding artifacts) as well as any artifacts resulting from the video compression which ensues. 

Ideally, we want to produce and work with dense importance maps annotated by real humans \emph{directly} on the compressed videos. These capture exactly the quantity we need, namely perceptual quality as it is jointly affected by \emph{everything}: the semantic content of the video, motion, video compression artifacts, and so on. To our knowledge, however, a method to collect such data has yet to be devised.

We propose a framework to collect and make use of fine-grained subjective importance information towards perceptual video compression. Our main contributions are:

\begin{enumerate}
    \item \textbf{Importance map collection tool (\pimtool{})}: We implemented a data collection web-tool which allows annotators to interactively construct a spatio-temporal perceptual importance map over the encoded videos, while keeping the bitrate of the video constant. 
    \item \textbf{Perceptual Importance Map Dataset (\pimd{})}: We use the tool to collect a dataset of human-annotated spatio-temporal importance maps over 178 diverse videos, totalling in 14443 densely annotated frames. We corroborate via subjective studies that encoding these videos based on importance maps, referred to as \textbf{\pimencoded{} videos}, indeed leads to better perceptual quality, with such videos preferred $1.8 \times$ more than baseline x264 encoded videos at the same bitrate.
    \item \textbf{Perceptual Importance Map Model (\pimm{})}: To showcase the utility of the curated dataset, we use it to train a simple CNN with interpretable features, to predict spatio-temporal importance maps for any video. We demonstrate via subjective studies that \textbf{predicted \pimencoded{} videos} yield improved perceptual quality for the same bitrate for a small test set, with predicted \pimencoded{} videos preferred $2 \times$ over baseline encoded videos.
\end{enumerate}

% We plan on releasing \pimtool{} and \pimd{}, along with associated code, upon acceptance.

\section{Related Work}
\label{sec:rel_work}
The common denominator across different strategies to video encoding is that the video is ultimately consumed by humans. As such, a considerable body of literature over the past two decades has been devoted to designing video encodings with particular appeal to the human visual system. As pointed out in reviews \cite{challenges_perceptual_compression, lee2012perceptual, zhang2021survey}, understanding of human perception can be incorporated into video compression across various levels: (1) improved \textit{perceptual modelling} of which parts of a video are important to a human viewer, (2) the \textit{implementation of perceptual coding} to incorporate this understanding into the underlying codec, and (3) the \textit{design of perceptual quality metrics} for evaluation.

\subsection{Perceptual Modeling}
\label{sec_rel_work_per_model}
The factors of perceptual importance to humans in videos are typically investigated via two main approaches: 
\begin{itemize}
    \item via \emph{low-level features}, such as contrast sensitivity \cite{nadenau2003wavelet, winkler1999perceptual}, luminance masking \cite{breitmeyer2006visual, enns2000s} and foveated compression \cite{itti2004automatic, wang2003foveation}, which can then can be incorporated to exploit visual redundancy in video coding.
    \item via \emph{high-level features}, such as faces or text, which account for human visual saliency. Remarkably, different humans can find different regions of video salient in a scene depending on the context. Human visual saliency data is often collected by providing users with a high-level task \cite{liu2010learning, zhang2018deep, zhang2017learning, wang2017learning}, or alternatively under natural attention settings \cite{Itti-saliency-based, cheng2014global, li2014secrets, lei2016universal, shi2015hierarchical}.
\end{itemize}

Saliency is often estimated via gaze-fixation \cite{judd2009learning, wilming2017extensive, borji2015cat2000, fan2018emotional}, or via salient object detection \cite{davsod, jiang2011automatic, hou2017deeply, cheng2014global, wang2017learning}. In gaze-fixation, saliency of objects is tracked by collecting and predicting human gaze: at present, such approach is inherently non-scalable as it requires special equipment in a laboratory setting. In salient object detection, pixel-wise segmentation maps are collected: these either focus on only one or few salient objects instead of the whole scene, and do not provide insight as to relative importance between multiple salient objects. Moreover, importantly, these methods only explain saliency originating from the raw scene, but do not account for compression artifacts which themselves carry significant implications on saliency. Beyond the video content which might capture the human attention, under practical settings videos also undergo compression prior to their consumption at the destination endpoint: this adds a further wrinkle to the study of saliency, as compressed videos feature various characteristic artifacts --- such as blurring, ringing and blocking --- that often catch viewers' attention. Furthermore, the variety of visual artifacts originating from the compression of various types of contents is challenging to model perceptually using low-level human visual features. In our data collection procedure, we allow users to paint the importance maps over the compressed videos with variable levels of importance --- allowing us to collect fine-grained yet scalable dataset which further takes into account the effect of compression.

\subsection{Implementation of Perceptual Coding}
One method to pursue perceptual video coding is emphasis on regions of interest (RoI) --- namely, investing bitrate to encode some regions of the video at higher quality than others. RoI-based methods typically employ one of the perceptual methods as described in \cref{sec_rel_work_per_model} to identify areas in a video which catch human attention and then use various modifications to the encoder to enable encoding them \cite{Itti_visual_attention_bit_aloc, itti_mpeg_compression, visual_saliency_compression, saliency_aware_compression, mckp, foveated_video_compression, Li2011VisualAG, Hadizadeh2014SaliencyAwareVC, guo2009novel, lyudvichenko2017semiautomatic, wang2022pvc}. Not-surprisingly, RoI-based perceptual models based solely on the content of the original video, often do not generalize well once the video is compressed, as they do not take into account the eye-catching visual artifacts introduced during compression. Our work can be thought of as a RoI-based video coding method where the region of interest is collected directly over compressed videos in an unconstrained fashion.

Another line of work to incorporate perceptual compression during video encoding does so by optimizing objective perceptual metrics such as SSIM \cite{wang2004image}, MS-SSIM \cite{wang2003multiscale} or VMAF \cite{li2018vmaf} instead of PSNR, via in-loop modifications to the video codec \cite{wang2011ssim, ou2011ssim, huang2021beyond, luo2019vmaf, deng2020vmaf}. These methods implicitly account for human perception by the virtue of utilizing a perceptual distortion metric in rate-distortion optimization. However, these objective metrics are known to only loosely correlate with the human visual system, and don't capture many prominent visual artifacts such as banding \cite{tandon2021cambi} or temporal artifacts \cite{madhusudana2021high}.

Finally, more recently, using machine learning for end-to-end learned video compression has gotten a lot of attention \cite{Rippel2019LearnedVC, feng2021versatile, lin2020m, mentzer2022vct, Hu2020ImprovingDV, Lu2020ContentAA, Yang2020LearningFV, Agustsson2020ScaleSpaceFF, rippel2021elf}. Our approach is complementary to these techniques, as a learned method can directly use the RoI found by our model to optimize the rate-distortion performance by using a weighted distortion similar to this work.

\section{PIMTool: Perceptual Importance Map collection Tool}
\label{sec:annotation_tool}
\begin{figure*}[!htb]
\centering
\includegraphics[width=0.9\textwidth]{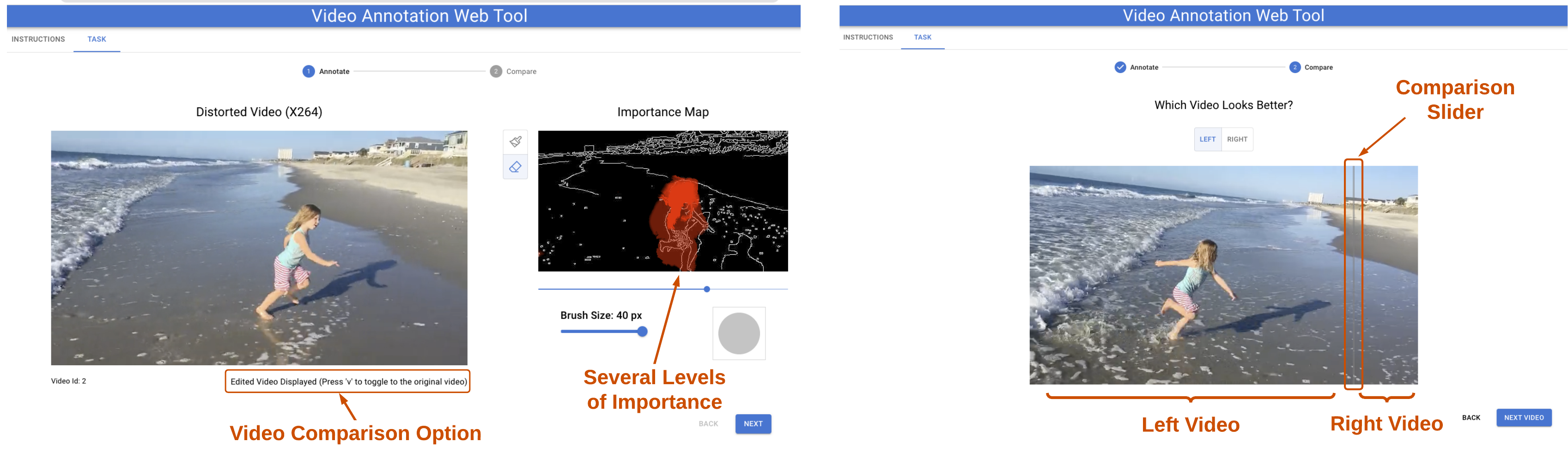}
\caption{An overview of \pimtool{} (Sec. \ref{sec:annotation_tool}). {\bf Left}: the main page visible to annotators during the collection of \pimd{}. {\bf Right}: the video comparison task which the annotators need to complete before their annotation is accepted.}
\label{sec3:fig:tool}
\end{figure*}

In \cite{our-tool} the authors developed an interactive web-tool which allows annotators to construct spatio-temporal \textit{perceptual importance maps} from a video, indicating areas they would want to see in a higher quality while keeping the encoding bitrate constant. Their tool provides real-time feedback to the annotator by iteratively re-encoding the original video based on the annotator-defined importance map, while maintaining the same bitrate. This allows annotator to visually understand the trade-off in encoding a region of the video at higher importance while degrading other regions because of the fixed-bitrate budget. 

Although this tool is a great initial prototype, it lacks several features necessary for large-scale data collection. Based on feedback from annotators, we made several modifications to the tool to enable such an effort. The changes include improvements to the video encoding procedure based on the importance map, improvements to the tool front-end enabling collection of higher accuracy importance maps,  and the addition of various quality control measures which are critical for collecting a large-scale high quality dataset.

\subsection{Improvements to the video encoding procedure}
\label{sec3:deltaQP}

At a high level, the x264 encoder works by cutting the video frame into \emph{macroblocks (\text{MB})} of size $16 \times 16$, and then encoding these macroblocks one at a time. The x264 encoder can trade off bits for quality mainly by modifying for each macroblock $\text{QP}_{\textrm{base}}$, the base \emph{quantization parameter} (QP) used to encode the video. The base QP parameter  takes values in the range $[0,51]$; higher values of base QP correspond to lower quality and vice versa. The choice of $\text{QP}_{\textrm{base}}$ used by x264 depends upon the target bitrate.

The x264 encoder can be used to adjust the quality in certain regions of the video by altering the QP value of the corresponding macroblock using a $\Delta \text{QP}$ map. The effective QP value used by x264 to encode a macroblock is just the sum of the $\text{QP}_{\textrm{base}}$ and the $\Delta \text{QP}$ value corresponding to the macroblock (i.e., $\text{QP}[\text{MB}] = \text{QP}_{\textrm{base}} + \Delta \text{QP}[\text{MB}]$). We can set the $\Delta$\text{QP} value for a macroblock to be negative when we want a higher quality (and vice versa). One limitation of the x264 encoder is that as the $\text{QP}[\text{MB}]$ increases beyond a certain value, the perceptual quality corresponding to that macroblock reduces dramatically due to the emergence of compression artifacts. Thus, care must be taken to keep the effective QP across the macroblocks from increasing beyond a certain value.  

The tool from \cite{our-tool} updates the displayed video by first translating the importance map to a $\Delta \text{QP}$ map and then using the $\Delta \text{QP}$ map during the video encoding, while maintaining the bitrate constant. This mapping was chosen to be linear in the range of $[-10, 10]$ irrespective of the fraction of the painted video. We found that this leads to videos with significant compression artifacts: if the $\Delta \text{QP}$ value is very low ($-10$ for example) in a large part of the video, the x264 encoder is forced to choose a higher $\text{QP}_{\textrm{base}}$ to maintain the average bitrate constant. A higher $\text{QP}_{\textrm{base}}$ in turn leads to a significantly higher effective QP in regions with high $\Delta \text{QP}$, resulting in significant compression artifacts due to the x264 encoding limitations described above. 

To mitigate these issues, instead of using a fixed range for $\Delta \text{QP}$, we adaptively modify this range based on the average value of the importance map, so that the bitrate change due to the $\Delta \text{QP}$ change is minimized. This is achieved by using a heuristic to estimate the bitrate change on changing the $\text{QP}[\text{MB}]$ of a macroblock, and then optimizing the range, to keep the overall bitrate unchanged.  We found that the heuristic $\text{bitrate[\textrm{\text{MB}}]}\propto 2^{\text{QP}[\text{MB}]/3}$, works well in practice. Using the adaptive QP-mapping leads to a significant reduction in cases with extreme unwanted distortions in re-encoded videos, and therefore better importance maps. 
We refer to videos re-encoded based on perceptual importance using this procedure as \textbf{\pimencoded{} videos}.

\subsection{Improvements to the tool front-end}
\label{sec3:tool:front-end}
Based on annotator feedback of the tool in \cite{our-tool}, we felt that there were certain design choices and missing features which were adversely affecting the annotator experience, and hence the quality of the collected importance maps. We describe these and how we addressed them below.

\paragraph{Ability to compare with the original video} In the original tool, the annotators would iteratively paint the importance maps and the tool would continually update the re-encoded video based on the cumulative importance map. However, there was no way for the annotator to actually compare their video reconstruction with the original baseline video. In PIMTool, we implemented a feature allowing the annotator to compare the original video to the video resulting from their painted maps, thus providing the annotator immediate and direct feedback (see \cref{sec3:fig:tool}).

\paragraph{Avoiding saturation in the importance maps} In the tool in \cite{our-tool}, the annotator can mark a spatio-temporal region of the video as important by clicking it. The longer the annotator clicks a region, the more important that region becomes. This makes the annotation more susceptible to saturation of the importance map. This unwanted saturation leads to large spatio-temporal regions being marked equally important, especially if a larger brush was used to color them. 

To overcome this, we take a two-step approach. In PIMTool, we allow painting the videos with only two fixed brush sizes --- a larger brush (40 pixels wide) allows painting coarse importance, and a smaller brush (20 pixels wide) allows painting the finer importance. To avoid saturation, the annotator can only increase importance in an area to 50\% using the larger brush, whereas they can increase the importance to 100\% using the smaller brush. This allows annotator to first coarsely paint important regions (e.g. whole body of a person) and then fine-tune by painting the most important regions using the smaller brush (e.g. the face of the person). We also instruct the annotators to paint approximately a third of the frame using the larger brush and a tenth using smaller brush to ensure that the annotator does not end up marking a large selection of video as important.

\subsection{Mechanisms for quality control}
For any crowd-sourced large-scale dataset collection, it is crucial to add means for quality control. In PIMTool, we added multiple ways to ensure that the quality of the annotated maps is on-par. One of the ways included checking if the annotator prefers the video they generated over the original baseline. After the annotator completes the task, they are presented with a comparison task to choose the better quality video between the final \pimencoded{} video and the original video. This is presented in the form of a slider (\cref{sec3:fig:tool}), and the video order is shuffled to avoid bias. We only accept the annotation if the annotator chooses the re-encoded video they created. We also collect more intermediate data in the PIMTool --- intermediate importance maps as the annotators are painting the video, intermediate QP-maps, and the answers to the video comparison. This allows us to perform better sanity checks, and better understand the video annotation process.

\section{PIMD: Perceptual Importance Map Dataset}
\label{sec:dataset}
\begin{figure}
\centering
\begin{subfigure}[t]{\columnwidth}
\includegraphics[width=\columnwidth]{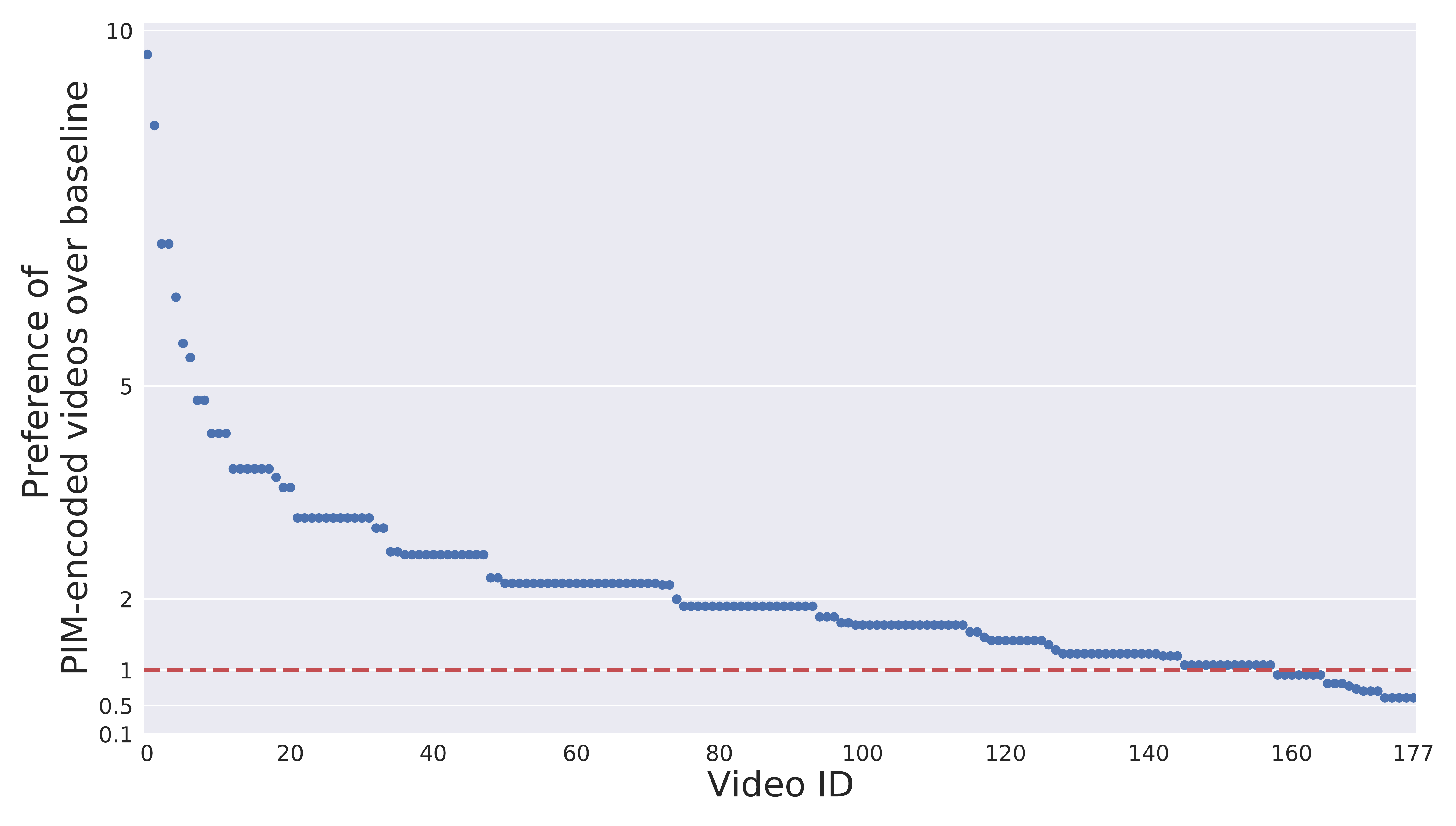}
\end{subfigure}
\caption{Subjective study results for \pimd{} (Sec. \ref{sec4:dataset:results}). For the 178 videos in \pimd{} we conducted a two-alternative forced choice (2AFC) subjective study with $\sim$30 raters between videos encoded with the collected importance maps versus the baseline encodings. The videos re-encoded using the collected importance map (\emph{\pimencoded{}} videos) were preferred on average 1.8$\times$ more than the baseline videos. The plot shows preferences across the entire dataset. The horizontal red line shows iso-preference.}
\label{sec4:fig:dataset_results}
\end{figure}

We used the \pimtool{} to collect a crowd-sourced spatio-temporal importance map dataset, which we call \pimd{} (\underline{P}erceptual \underline{I}mportance \underline{M}ap \underline{D}ataset). \pimd{} consists of
\begin{itemize}
    \item 178 {\bf videos} of resolution $450\times800$; and
    \item for each frame of each video, spatio-temporal 0-255 {\bf importance maps} of resolution $270 \times 480$
\end{itemize}
Visual examples from the dataset can be found in \cref{sec4:fig:dataset_example} and \cref{sec5:fig:examples}. Below we first elaborate on the dataset collection process, and provide quantitative and qualitative analyses.

\subsection{Data Collection}

\paragraph{Video Selection} We collected importance maps for videos from three different public datasets: 25 videos from CLIC2021 (subset of the UGC dataset\footnote{\url{https://storage.googleapis.com/clic2021_public/txt_files/video_urls.txt}}), 93 videos from BVI-DVC \cite{bvi-dvc}, and 60 videos from Pexels\footnote{\url{https://www.pexels.com}}. The videos are cropped to $450\times800$ as required by \pimtool{}, are 1-5s in duration, have no scene cuts, and play in a range of 25-120 FPS during the data collection process.

To promote diversity, each video in our dataset was annotated by all annotators from one of two groups. In the first group we had two expert annotators who are familiar with the compression artifacts, whereas in the second group there were five professional annotators who are not necessarily familiar with compression. Finally, individual annotations per video were averaged resulting in the average importance map which is used for further analysis in this work.

\paragraph{Video Bitrates} The utility of the importance maps for video encoding is particularly evident when the bitrate is limited, as it allows annotators to shift more bits to the areas they want to see in a higher quality from areas that are less important. In selecting the bitrate target, our objective was to choose values which are low but still realistic. As the bitrate for a video depends heavily on the complexity of the video (such as videos without much motion have low bitrates), we encode each video at a bitrate that matches a fixed VMAF \cite{li2018vmaf} quality value. 

We conducted a two-alternative forced choice
(2AFC) subjective pilot study to understand the trade-off between bitrate selection and encoding advantages from using the importance map based encoding. Based on the study results, we chose a bitrate that matches a score of VMAF 70 for our whole dataset. Please see the supplementary material for more details.

\subsection{Using importance maps from \pimd{} for perceptual encoding}
\label{sec4:dataset:results}

To validate the efficacy of the importance maps collected as part of \pimd{}, we analyzed the dataset qualitatively as well as quantitatively. We used importance maps collected in \pimd{} to create final \emph{\pimencoded{}} videos using the $\Delta \text{QP}$-based x264 encoding described in \cref{sec3:deltaQP}.

\subsubsection{Quantitative Analysis}
For each video in \pimd{}, the \pimencoded{} videos were compared against the baseline x264-encoded videos without any spatio-temporal importance, at a similar bitrate, in a subjective study involving $\sim$30 different observers using a two-alternative forced choice (2AFC) approach. In a 2AFC video comparison approach the viewers are asked to watch two videos and then they must select the preferred video. We collected the data through three surveys without any intersecting videos between the surveys. We recorded 29, 32, and 33 viewer responses for these three surveys. We calculate the preference over the dataset by first calculating the fraction of users who prefer \pimencoded{} videos over the baseline videos (say $p$), and then we convert it to the preference ($p/(1-p)$). The \pimencoded{} videos were preferred 1.8$\times$ over the baseline videos, with 158 out of 178 of the videos ($\approx88\%$) preferred over the baseline.  Similarly, for each video in the dataset, let $q_i$  be the fraction of viewers who prefer the  $i^{th}$ \pimencoded{} video then the preference for the $i^{th}$ video is defined by $q_i/(1-q_i)$. \cref{sec4:fig:dataset_results} shows the distribution of preferences over the complete dataset. 

\subsubsection{Qualitative Analysis}
\label{sec4:dataset:results:qualitative}
\cref{sec4:fig:dataset_example} shows a typical example from \pimd{} highlighting the trade-offs in bitrate and perceptual quality. The video from which this frame was extracted was preferred 6.25$\times$ over the baseline video in our subjective study. The importance map shows that the video has roughly three annotation categories: not important, medium important and very important. The face of the person is the most visually important region in this video and the x264 encoding based on the collected importance map improves the face of the person significantly (blue bounding box).

Interestingly, saliency-based models would choose the whole person or the text as equally important features in this video --- but as can been seen in the collected data, the body of the person is not as visually important as the face of the person and hence is not worth trading as many bits for. \pimd{} captures such effects as can be seen in \cref{sec4:fig:dataset_example} green bounding box: even though the quality of the body of the person is improved, it is not as significant. Similarly, even though text is considered a salient feature, in this example annotators unanimously prioritized the person over the text. But as can be seen in \cref{sec4:fig:dataset_example} red bounding box, this results in a slight degradation of the text quality in the \pimencoded{} video. Within this example, we see that allocating bits away from the text and into the person's face is much more beneficial towards perceptual quality --- although in other contexts the text might be more important. \pimtool{} and \pimd{} allow teasing out such subtleties in the videos, which can be exploited for better encoding.

\section{PIMM: Perceptual Importance Map Model}
\label{sec:model}
\begin{figure*}
\centering
\begin{subfigure}[t]{\textwidth}
\includegraphics[width=\textwidth]{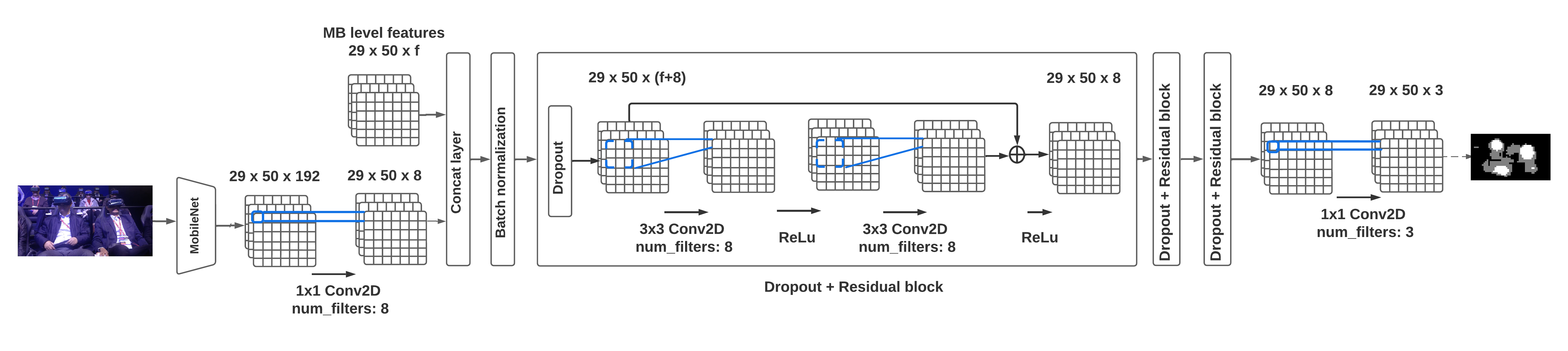}
\end{subfigure}
\caption{\pimm{} Architecture. \pimm{} predicts the importance maps for unseen videos by combining multiple simple and interpretable features such as frame-level object saliency and segmentation embeddings capturing high-level information, with macroblock-level distortion features from quality metrics such as PSNR, SSIM, VIF capturing low-level information. See \cref{sec5:model:description} for more details.}
\label{sec5:fig:model}
\end{figure*}

\begin{figure}
\centering
\begin{subfigure}[t]{\columnwidth}
\includegraphics[width=\textwidth]{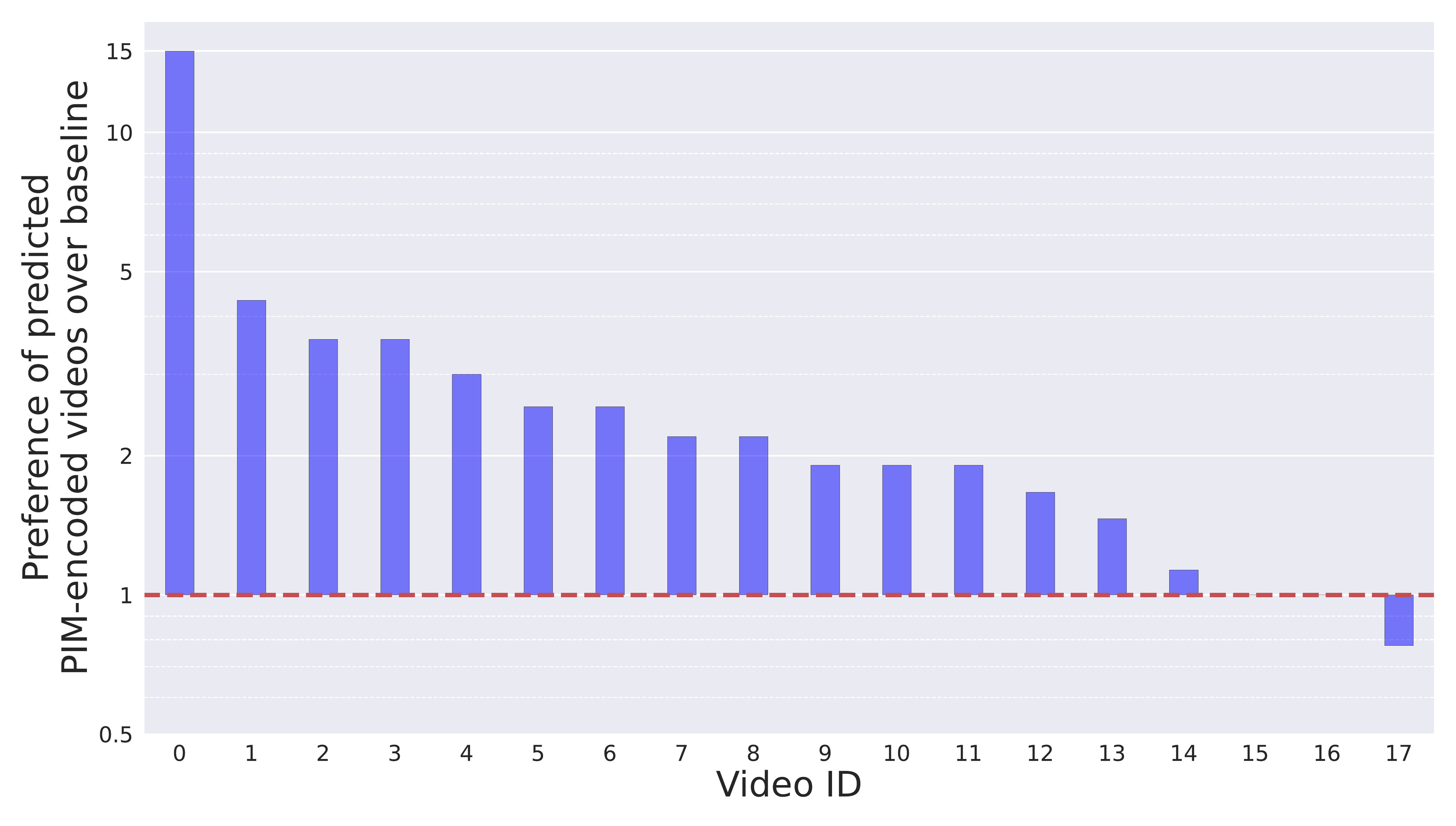}
\end{subfigure}
\caption{Subjective study results for our importance map prediction model. For the 18 videos in \pimd{} test set we conducted another 2AFC subjective study (with 32 raters) between videos encoded with the predicted importance maps (\emph{predicted \pimencoded{}}) videos and the baseline videos. The predicted \pimencoded{} videos were preferred on average 2$\times$ over the baseline videos.}
\label{sec5:fig:model_test-set}
\end{figure}

\begin{figure*}
\centering
\includegraphics[width=0.9\textwidth]{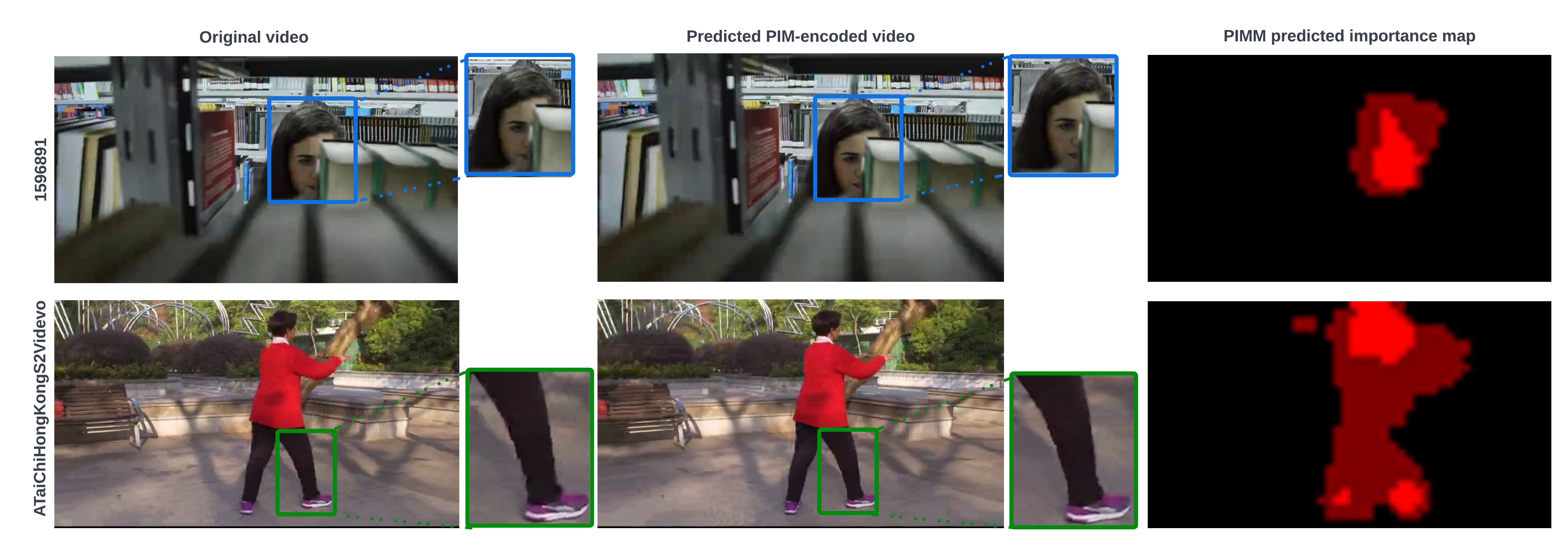}
\caption{Examples of importance map predictions by \pimm{} (\cref{sec5:model:results}). For each video, we see that \pimm{} predicted importance maps which capture salient regions in the video lead to perceptually-improved videos over the baseline.}
\label{sec5:fig:examples}
\end{figure*}

As was shown in \cref{sec:dataset}, \pimencoded{} videos from \pimd{} are preferred by observers over the baseline videos encoded without importance maps. However, collecting importance maps with humans in the loop is clearly impractical as it is time consuming and expensive. Instead, we show that a lightweight CNN with simple, interpretable features can be trained on \pimd{} to predict the human-generated importance maps. We refer to the model as a {\underline P}erceptual {\underline I}mportance {\underline M}ap {\underline M}odel (\pimm{}). In this section, we present \pimm{}'s architecture, training configuration and results. 

\subsection{Model Overview}
\label{sec5:model:description}
\paragraph{Problem setup} We trained a three-class classification model that operates per-frame and predicts the importance levels of each macroblock. The three classes can be thought of as different levels of importance, mapped to importance values 0, 128 and 255 respectively. This setup is motivated by the tri-modal importance distribution observed in \pimd{} (see \cref{sec3:tool:front-end} and \cref{sec4:dataset:results:qualitative} for more details). These maps are converted to ${\Delta \text{QP}}$ maps in the same way as in the \pimtool{}, which are then used for encoding using x264. The videos re-encoded using predicted importance maps are referred to as \textbf{predicted \pimencoded{} videos}. To avoid overfitting due to the limited number of videos and the high temporal correlation between frames, we train a lightweight CNN which contains $\sim$8K trainable parameters.

\paragraph{Model architecture}
The model consists of three consecutive residual blocks, whose convolutions have 8 filters each and kernels of size $3 \times 3$. The first residual block also has an additional convolution along the skip-connection to match the output shape. To mitigate overfitting, a dropout layer is added before each residual block. Finally we have a 3-channel $1 \times 1$ convolution which produces the scores for the three classes, as shown in  \cref{sec5:fig:model}.

The inputs to the model comprise a variety of useful features --- both high- and low-level --- each of which provides a different signal indicative of the resultant perceptual importance map. Each feature was validated to provide a meaningful contribution to the final prediction (see supplementary material for ablation studies). All the features are rescaled to dimension $29 \times 50$ (that is, to produce one value per MB) and are concatenated together. The model returns softmax scores of dimension $29 \times 50 \times 3$, which are then used to predict the importance level assigned to each MB. 

The input features to the model include:

\begin{itemize}
    \item \textit{Object segmentation map} is computed using a pre-trained DeepLabv3 model \cite{deeplabv3}, which outputs a segmentation map with 21 classes, which we feed to the model in its one-hot representation.
    \item \textit{Saliency map} is computed using \textit{SSAV} model \cite{davsod,ssav}.
    \item \textit{MobileNetV2 embeddings} are computed using a frozen pre-trained MobileNetV2 \cite{mobilenetv2}. Specifically, we use the output of \textit{block\_6\_depthwise} layer. 
    \item \textit{Optical flow} is computed using Farneback algorithm (OpenCV implementation)\footnote{OpenCV Farneback implementation: \url{https://docs.opencv.org/3.4/d4/dee/tutorial_optical_flow.html}}.
    
    \item \textit{PSNR} and \textit{SSIM} values are computed on the luminosity channel of the frame for each macroblock (size $16 \times 16$), creating a single-channel input feature each. 
    \item \textit{VIF}\cite{vif_metric} is computed similarly to the other quality metrics for each macroblock, but at a block size of $256 \times 256$ and a stride of $16$. 
    \item \textit{Spatial high pass filter} takes the difference between a pixel and the average value of the surrounding pixels for each color channel of the encoded frame.
    \item \textit{Temporal high pass filter} is computed similarly to the spatial high pass filter, but takes the difference between pixels of frame ${i}$ and the average value of pixels from frames ${i+1}$ and ${i-1}$.
\end{itemize}

More details on the choice of features is described as a part of the model ablation studies in the supplementary material section. Because different features have different magnitude scales, a batch normalization layer is added after all the inputs are concatenated.

\paragraph{Implementation and Training} We use the \pimd{} dataset (\cref{sec:dataset}) for training the model. We randomly partition the dataset into train, validation and test sets consisting of 143, 17 and 18 videos respectively. We use standard data augmentation techniques such as horizontal/vertical flipping to artificially increase the size of the training set. The model is trained on an NVIDIA Titan X GPU for 70 epochs (499 iterations per epoch) using weighted cross-entropy loss with the Adam optimizer and an initial learning rate of $10^{-4}$. Please see the supplementary material for more details.

\begin{figure}[b]
\centering
\begin{subfigure}[t]{\columnwidth}
\includegraphics[width=\columnwidth]{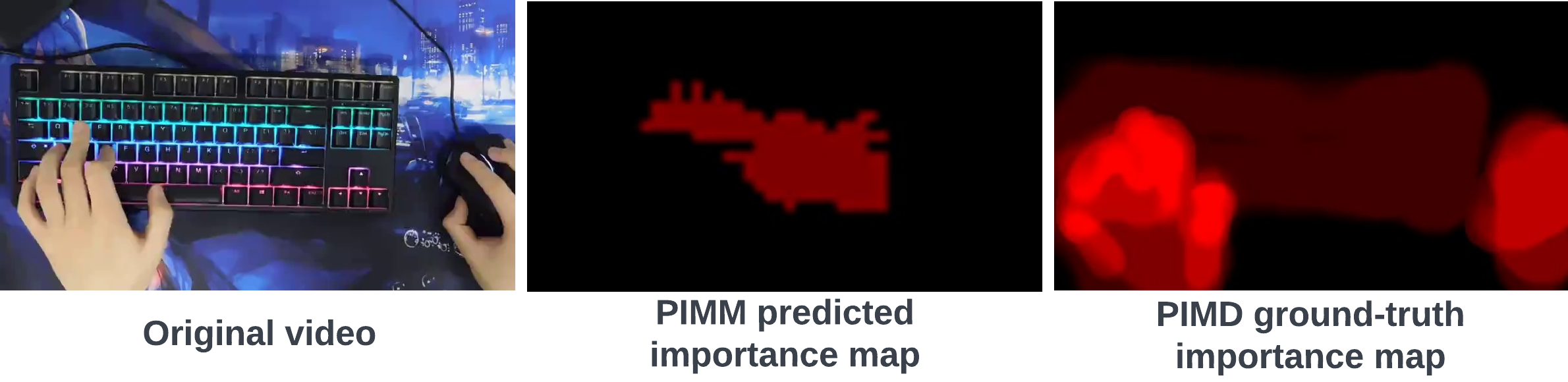}
\end{subfigure}
\caption{PIMM failure case (\cref{sec5:model:results}). In this video from \pimd{} test set, the \pimm{} predicted map differs significantly from the ground truth importance map in \pimd{} leading to the predicted \pimencoded{} video underperforming the baseline.}
\label{sec5:fig:negative_example}
\end{figure}

\subsection{Importance Map Prediction Results}
\label{sec5:model:results}

To assess the performance of \pimm{}, we evaluate it on \pimd{} test set consisting of 18 videos to obtain predicted importance maps for each video. The predicted \pimencoded{} videos are compared to baseline videos in another subjective study with different observers ($n=32$) as described in \cref{sec4:dataset:results}. 

\paragraph{Subjective study results} As shown in \cref{sec5:fig:model_test-set}, the predicted \pimencoded{} videos are more than $2 \times$ likely to be preferred by observers compared to the baseline, with 15/18 videos preferred over the baseline. Comparatively, the test set ground-truth \pimencoded{} videos were preferred $2.8 \times$.

\paragraph{Bitrate savings at the same perceptual quality} In order to understand the bitrate savings that can be achieved by using the PIM-based encoding, we repeated the subjective study with the baseline x264 encoded videos encoded at $1.05\times$, $1.1\times $ and $ 1.2\times$ the bitrate at VMAF 70. The subjective study suggested that predicted \pimencoded{} videos were preferred over these baselines at factors $1.22 \times $, $1.1 \times $ and $0.95 \times$ respectively. Linearly interpolating these results, we estimate that using the predicted \pimencoded{} videos led to an average bitrate savings of $\sim 15\%$ over the baseline encodes at the same perceptual quality. 

\paragraph{Qualitative analysis} \cref{sec5:fig:examples} shows two examples of predicted \pimencoded{} videos. The predicted \pimencoded{} video \emph{159689} is $15 \times$ more likely to be chosen compared to the baseline. The predicted map of this video is very similar to the average importance map from \pimd{}, leading to a significant quality improvement in the girl's face. The predicted \pimencoded{} video \emph{ATaiChiHongKongS2Videvo} is preferred $2.2 \times$ over the baseline. For this video, the predicted importance map focuses on the body of the person, assigning higher importance to the feet and head. This results in a smoother video without distracting artifacts, leading to observers preferring this video over the baseline.

Although videos encoded using our method are twice as likely to be preferred compared to the baseline, there might be videos, such as \emph{979689} in  \cref{sec5:fig:negative_example}, where we slightly underperform compared to the baseline. Our model's predicted map does not correctly predict the important areas. Because the bitrate is limited, improving the quality in certain areas inevitably means decreasing the quality in other areas. While our tool limits the $\Delta \text{QP}$ values (\cref{sec:annotation_tool}), and the decrease in quality should not be substantial, there are cases where they might be noticeable.

\paragraph{Using the importance maps to encode at other bitrates} The \pimd{} importance maps were collected at bitrates corresponding to VMAF score of 70, and as such \pimm{} learned to predict importance maps corresponding to this VMAF value. It is however interesting to explore whether the \pimm{}-predicted importance maps generalize beyond the bitrates at which they were trained. To understand this, we used the predicted importance maps from \pimm{} but use them to encode the test set videos at bitrates matching VMAF 65 and VMAF 75. We then repeated the subjective study as described in \cref{sec4:dataset:results}.

For the VMAF 75 comparisons, the predicted \pimencoded{} videos were preferred on average $1.3 \times$ over the baseline. This drop in performance, compared to VMAF 70 is somewhat expected, as an increase in the bitrate leads to better overall quality of the baseline x264 video, and hence less room for improvement. As for the VMAF 65 comparison, the predicted \pimencoded{} videos were preferred on average $1.78 \times$ over the baseline. This slight drop in performance suggests that though the generated importance maps are indeed specialized to the bitrates corresponding to VMAF 70, they still provide significant benefits at bitrates corresponding to nearby VMAF. One reason for this specialization is that the observed compression distortions change based on the bitrate, thus requiring slightly different importance maps at different bitrates.

\section{Discussions and Conclusion}
\label{sec:discussion}
In this work, we described a framework for collecting and using fine-grained subjective importance maps towards perceptual video compression. We developed an annotation tool, and used it to collect a dataset containing fine-grained importance maps. We corroborated through a subjective study that videos re-encoded using collected importance maps, referred to as \pimencoded{} videos, are preferred $1.8 \times$ over the baseline encoded videos at the same bitrate (using x264). We also demonstrated that the dataset can be used to train a small proof-of-concept model to predict importance maps for unseen videos, and re-encoding using these predicted maps produces videos that are preferred $2 \times$ at the same bitrate when evaluated on a small test set.

We observed that the preference for the ground-truth \pimencoded{} test set videos was $2.8 \times$ (\cref{sec5:model:results}), while the preference of \pimencoded{} videos across the entire \pimd{} was $1.8\times$ (\cref{sec4:dataset:results}). We believe that the difference in the preference ratios between the test set and the entire \pimd{} is primarily due to the fact that the ground-truth results for the test set came entirely from one of the three surveys (\cref{sec4:dataset:results}). To elaborate, \pimd{} was randomly split into train, val, and test sets before the ground-truth data was collected. Therefore, we hypothesize that the viewers in the survey from which the test ground truth data was collected might have had a slight preference for \pimencoded{} videos. We believe that this difference can be minimized by expanding the size of \pimd{} and evaluating the \pimencoded{} videos with a larger group of viewers.

While this work builds an initial framework for fine-grained importance, significant opportunities remain to further improve it. For example, while \pimd{} is a good starting point, it can certainly be expanded to a larger set of videos. \pimtool{} should facilitate convenient collection of a larger dataset at different bitrates, and can be generalized to other codec specifications such as H.265, AV1 and even ML-based learned video codecs. One interesting question here would be to study whether importance maps collected using different codecs correlate. Another interesting line of future work is improving \pimm{} to predict higher quality importance maps. We chose the current framework to keep the model size small, given the moderate size of \pimd{}. With a larger dataset, models with higher number of parameters and therefore potentially better models could be trained.

\section{Acknowledgements}
\label{sec:ack}
We would like to thank Eric Sun for his help with the annotations.

%%%%%%%%% REFERENCES
{\small
\bibliographystyle{ieee_fullname}
\bibliography{egbib}

\begin{thebibliography}{10}\itemsep=-1pt

\bibitem{Agustsson2020ScaleSpaceFF}
Eirikur Agustsson, David~C. Minnen, Nick Johnston, Johannes Ball{\'e}, Sung~Jin
  Hwang, and George Toderici.
\newblock Scale-space flow for end-to-end optimized video compression.
\newblock {\em 2020 IEEE/CVF Conference on Computer Vision and Pattern
  Recognition (CVPR)}, pages 8500--8509, 2020.

\bibitem{borji2015cat2000}
Ali Borji and Laurent Itti.
\newblock {Cat2000: A large scale fixation dataset for boosting saliency
  research}.
\newblock {\em arXiv preprint arXiv:1505.03581}, 2015.

\bibitem{breitmeyer2006visual}
Bruno Breitmeyer, Haluk Ogmen, Haluk {\"O}{\u{g}}men, et~al.
\newblock {\em Visual masking: Time slices through conscious and unconscious
  vision}.
\newblock Oxford University Press, 2006.

\bibitem{deeplabv3}
Liang{-}Chieh Chen, George Papandreou, Florian Schroff, and Hartwig Adam.
\newblock Rethinking atrous convolution for semantic image segmentation.
\newblock {\em CoRR}, abs/1706.05587, 2017.

\bibitem{challenges_perceptual_compression}
Zhenzhong Chen, Weisi Lin, and King~Ngi Ngan.
\newblock Perceptual video coding: Challenges and approaches.
\newblock In {\em 2010 IEEE International Conference on Multimedia and Expo},
  pages 784--789, 2010.

\bibitem{cheng2014global}
Ming-Ming Cheng, Niloy~J Mitra, Xiaolei Huang, Philip~HS Torr, and Shi-Min Hu.
\newblock Global contrast based salient region detection.
\newblock {\em IEEE transactions on pattern analysis and machine intelligence},
  37(3):569--582, 2014.

\bibitem{deng2020vmaf}
Sai Deng, Jingning Han, and Yaowu Xu.
\newblock Vmaf based rate-distortion optimization for video coding.
\newblock In {\em 2020 IEEE 22nd International Workshop on Multimedia Signal
  Processing (MMSP)}, pages 1--6. IEEE, 2020.

\bibitem{itti_mpeg_compression}
N. Dhavale and L. Itti.
\newblock Saliency-based multifoveated mpeg compression.
\newblock In {\em Seventh International Symposium on Signal Processing and Its
  Applications, 2003. Proceedings.}, volume~1, pages 229--232 vol.1, 2003.

\bibitem{enns2000s}
James~T Enns and Vincent Di~Lollo.
\newblock What’s new in visual masking?
\newblock {\em Trends in cognitive sciences}, 4(9):345--352, 2000.

\bibitem{davsod}
Deng-Ping Fan, Wenguan Wang, Ming-Ming Cheng, and Jianbing Shen.
\newblock Shifting more attention to video salient object detection.
\newblock In {\em IEEE CVPR}, 2019.

\bibitem{fan2018emotional}
Shaojing Fan, Zhiqi Shen, Ming Jiang, Bryan~L Koenig, Juan Xu, Mohan~S
  Kankanhalli, and Qi Zhao.
\newblock {Emotional attention: A study of image sentiment and visual
  attention}.
\newblock In {\em Proceedings of the IEEE Conference on computer vision and
  pattern recognition}, pages 7521--7531, 2018.

\bibitem{feng2021versatile}
Runsen Feng, Zongyu Guo, Zhizheng Zhang, and Zhibo Chen.
\newblock Versatile learned video compression.
\newblock {\em arXiv preprint arXiv:2111.03386}, 2021.

\bibitem{guo2009novel}
Chenlei Guo and Liming Zhang.
\newblock A novel multiresolution spatiotemporal saliency detection model and
  its applications in image and video compression.
\newblock {\em IEEE transactions on image processing}, 19(1):185--198, 2009.

\bibitem{visual_saliency_compression}
Rupesh Gupta, Meera {Thapar Khanna}, and Santanu Chaudhury.
\newblock Visual saliency guided video compression algorithm.
\newblock {\em Signal Processing: Image Communication}, 28(9):1006--1022, 2013.

\bibitem{Hadizadeh2014SaliencyAwareVC}
Hadi Hadizadeh and Ivan~V. Baji{\'c}.
\newblock Saliency-aware video compression.
\newblock {\em IEEE Transactions on Image Processing}, 23:19--33, 2014.

\bibitem{mckp}
Hadi Hadizadeh and Ivan~V. Bajić.
\newblock Saliency-preserving video compression.
\newblock In {\em 2011 IEEE International Conference on Multimedia and Expo},
  pages 1--6, 2011.

\bibitem{saliency_aware_compression}
Hadi Hadizadeh and Ivan~V. Bajić.
\newblock Saliency-aware video compression.
\newblock {\em IEEE Transactions on Image Processing}, 23(1):19--33, 2014.

\bibitem{hou2017deeply}
Qibin Hou, Ming-Ming Cheng, Xiaowei Hu, Ali Borji, Zhuowen Tu, and Philip~HS
  Torr.
\newblock Deeply supervised salient object detection with short connections.
\newblock In {\em Proceedings of the IEEE conference on computer vision and
  pattern recognition}, pages 3203--3212, 2017.

\bibitem{Hu2020ImprovingDV}
Zhihao Hu, Zhenghao Chen, Dong Xu, Guo Lu, Wanli Ouyang, Shuhang Gu~College of
  Software, Beihang University, China., School of Electrical, Information
  Engineering, The~University of Sydney, Australia, School of
  Computer~ScienceTechnology, and Beijing~University of Technology.
\newblock Improving deep video compression by resolution-adaptive flow coding.
\newblock In {\em ECCV}, 2020.

\bibitem{huang2021beyond}
Zhimeng Huang, Kai Lin, Chuanmin Jia, Shanshe Wang, and Siwei Ma.
\newblock Beyond vvc: Towards perceptual quality optimized video compression
  using multi-scale hybrid approaches.
\newblock In {\em Proceedings of the IEEE/CVF Conference on Computer Vision and
  Pattern Recognition}, pages 1866--1869, 2021.

\bibitem{itti2004automatic}
Laurent Itti.
\newblock Automatic foveation for video compression using a neurobiological
  model of visual attention.
\newblock {\em IEEE transactions on image processing}, 13(10):1304--1318, 2004.

\bibitem{foveated_video_compression}
L. Itti.
\newblock Automatic foveation for video compression using a neurobiological
  model of visual attention.
\newblock {\em IEEE Transactions on Image Processing}, 13(10):1304--1318, 2004.

\bibitem{Itti-saliency-based}
L. Itti, C. Koch, and E. Niebur.
\newblock A model of saliency-based visual attention for rapid scene analysis.
\newblock {\em IEEE Transactions on Pattern Analysis and Machine Intelligence},
  20(11):1254--1259, 1998.

\bibitem{jiang2011automatic}
Huaizu Jiang, Jingdong Wang, Zejian Yuan, Tie Liu, Nanning Zheng, and Shipeng
  Li.
\newblock Automatic salient object segmentation based on context and shape
  prior.
\newblock In {\em BMVC}, volume~6, page~9, 2011.

\bibitem{judd2009learning}
Tilke Judd, Krista Ehinger, Fr{\'e}do Durand, and Antonio Torralba.
\newblock Learning to predict where humans look.
\newblock In {\em 2009 IEEE 12th international conference on computer vision},
  pages 2106--2113. IEEE, 2009.

\bibitem{lee2012perceptual}
Jong-Seok Lee and Touradj Ebrahimi.
\newblock Perceptual video compression: A survey.
\newblock {\em IEEE Journal of selected topics in signal processing},
  6(6):684--697, 2012.

\bibitem{lei2016universal}
Jianjun Lei, Bingren Wang, Yuming Fang, Weisi Lin, Patrick Le~Callet, Nam Ling,
  and Chunping Hou.
\newblock A universal framework for salient object detection.
\newblock {\em IEEE Transactions on Multimedia}, 18(9):1783--1795, 2016.

\bibitem{li2014secrets}
Yin Li, Xiaodi Hou, Christof Koch, James~M Rehg, and Alan~L Yuille.
\newblock The secrets of salient object segmentation.
\newblock In {\em Proceedings of the IEEE conference on computer vision and
  pattern recognition}, pages 280--287, 2014.

\bibitem{li2018vmaf}
Zhi Li, Christos Bampis, Julie Novak, Anne Aaron, Kyle Swanson, Anush Moorthy,
  and JD Cock.
\newblock Vmaf: The journey continues.
\newblock {\em Netflix Technology Blog}, 25, 2018.

\bibitem{Itti_visual_attention_bit_aloc}
Zhicheng Li, Shiyin Qin, and Laurent Itti.
\newblock Visual attention guided bit allocation in video compression.
\newblock {\em Image and Vision Computing}, 29(1):1--14, 2011.

\bibitem{Li2011VisualAG}
Zhicheng Li, Shiyin Qin, and Laurent Itti.
\newblock Visual attention guided bit allocation in video compression.
\newblock {\em Image Vis. Comput.}, 29:1--14, 2011.

\bibitem{lin2020m}
Jianping Lin, Dong Liu, Houqiang Li, and Feng Wu.
\newblock M-lvc: Multiple frames prediction for learned video compression.
\newblock In {\em Proceedings of the IEEE/CVF Conference on Computer Vision and
  Pattern Recognition}, pages 3546--3554, 2020.

\bibitem{liu2010learning}
Tie Liu, Zejian Yuan, Jian Sun, Jingdong Wang, Nanning Zheng, Xiaoou Tang, and
  Heung-Yeung Shum.
\newblock Learning to detect a salient object.
\newblock {\em IEEE Transactions on Pattern analysis and machine intelligence},
  33(2):353--367, 2010.

\bibitem{Lu2020ContentAA}
Guo Lu, Chunlei Cai, Xiaoyun Zhang, Li Chen, Wanli Ouyang, Dong Xu, and Zhiyong
  Gao.
\newblock Content adaptive and error propagation aware deep video compression.
\newblock In {\em ECCV}, 2020.

\bibitem{luo2019vmaf}
Zhengyi Luo, Yan Huang, Xiangwen Wang, Rong Xie, and Li Song.
\newblock Vmaf oriented perceptual optimization for video coding.
\newblock In {\em 2019 IEEE International Symposium on Circuits and Systems
  (ISCAS)}, pages 1--5. IEEE, 2019.

\bibitem{lyudvichenko2017semiautomatic}
Vitaliy Lyudvichenko, Mikhail Erofeev, Yury Gitman, and Dmitriy Vatolin.
\newblock A semiautomatic saliency model and its application to video
  compression.
\newblock In {\em 2017 13th IEEE International Conference on Intelligent
  Computer Communication and Processing (ICCP)}, pages 403--410. IEEE, 2017.

\bibitem{bvi-dvc}
Di Ma, Fan Zhang, and David~R. Bull.
\newblock {BVI}-{DVC}: A training database for deep video compression.
\newblock {\em {IEEE} Transactions on Multimedia}, 24:3847--3858, 2022.

\bibitem{madhusudana2021high}
Pavan~C Madhusudana, Neil Birkbeck, Yilin Wang, Balu Adsumilli, and Alan~C
  Bovik.
\newblock High frame rate video quality assessment using vmaf and entropic
  differences.
\newblock In {\em 2021 Picture Coding Symposium (PCS)}, pages 1--5. IEEE, 2021.

\bibitem{mentzer2022vct}
Fabian Mentzer, George Toderici, David Minnen, Sung-Jin Hwang, Sergi Caelles,
  Mario Lucic, and Eirikur Agustsson.
\newblock Vct: A video compression transformer.
\newblock {\em arXiv preprint arXiv:2206.07307}, 2022.

\bibitem{nadenau2003wavelet}
Marcus~J Nadenau, Julien Reichel, and Murat Kunt.
\newblock Wavelet-based color image compression: exploiting the contrast
  sensitivity function.
\newblock {\em IEEE Transactions on image processing}, 12(1):58--70, 2003.

\bibitem{ou2011ssim}
Tao-Sheng Ou, Yi-Hsin Huang, and Homer~H Chen.
\newblock Ssim-based perceptual rate control for video coding.
\newblock {\em IEEE Transactions on Circuits and Systems for Video Technology},
  21(5):682--691, 2011.

\bibitem{our-tool}
Evgenya Pergament, Pulkit Tandon, Kedar Tatwawadi, Oren Rippel, Lubomir~D.
  Bourdev, Bruno~A. Olshausen, Tsachy Weissman, Sachin Katti, and Alexander~G.
  Anderson.
\newblock An interactive annotation tool for perceptual video compression.
\newblock In {\em 14th International Conference on Quality of Multimedia
  Experience, QoMEX 2022, Lippstadt, Germany, September 5-7, 2022}, pages 1--6.
  {IEEE}, 2022.

\bibitem{rippel2021elf}
Oren Rippel, Alexander~G Anderson, Kedar Tatwawadi, Sanjay Nair, Craig Lytle,
  and Lubomir Bourdev.
\newblock Elf-vc: Efficient learned flexible-rate video coding.
\newblock In {\em Proceedings of the IEEE/CVF International Conference on
  Computer Vision}, pages 14479--14488, 2021.

\bibitem{Rippel2019LearnedVC}
Oren Rippel, Sanjay Nair, Carissa Lew, Steve Branson, Alexander~G. Anderson,
  and Lubomir~D. Bourdev.
\newblock Learned video compression.
\newblock {\em 2019 IEEE/CVF International Conference on Computer Vision
  (ICCV)}, pages 3453--3462, 2019.

\bibitem{mobilenetv2}
Mark Sandler, Andrew Howard, Menglong Zhu, Andrey Zhmoginov, and Liang-Chieh
  Chen.
\newblock Mobilenetv2: Inverted residuals and linear bottlenecks.
\newblock 2018.

\bibitem{vif_metric}
H.R. Sheikh and A.C. Bovik.
\newblock Image information and visual quality.
\newblock {\em IEEE Transactions on Image Processing}, 15(2):430--444, 2006.

\bibitem{shi2015hierarchical}
Jianping Shi, Qiong Yan, Li Xu, and Jiaya Jia.
\newblock {Hierarchical image saliency detection on extended CSSD}.
\newblock {\em IEEE transactions on pattern analysis and machine intelligence},
  38(4):717--729, 2015.

\bibitem{ssav}
Hongmei Song, Wenguan Wang, Sanyuan Zhao, Jianbing Shen, and Kin-Man Lam.
\newblock Pyramid dilated deeper convlstm for video salient object detection.
\newblock In {\em ECCV}, pages 715--731, 2018.

\bibitem{tandon2021cambi}
Pulkit Tandon, Mariana Afonso, Joel Sole, and Luk{\'a}{\v{s}} Krasula.
\newblock Cambi: Contrast-aware multiscale banding index.
\newblock In {\em 2021 Picture Coding Symposium (PCS)}, pages 1--5. IEEE, 2021.

\bibitem{wang2022pvc}
Bing Wang, Pan Gao, Qiang Peng, Qianwen Zhang, Xiao Wu, and Wei Xiang.
\newblock Pvc-stim: Perceptual video coding based on spatio-temporal influence
  map.
\newblock {\em Signal, Image and Video Processing}, pages 1--9, 2022.

\bibitem{wang2017learning}
Lijun Wang, Huchuan Lu, Yifan Wang, Mengyang Feng, Dong Wang, Baocai Yin, and
  Xiang Ruan.
\newblock Learning to detect salient objects with image-level supervision.
\newblock In {\em Proceedings of the IEEE conference on computer vision and
  pattern recognition}, pages 136--145, 2017.

\bibitem{wang2011ssim}
Shiqi Wang, Abdul Rehman, Zhou Wang, Siwei Ma, and Wen Gao.
\newblock Ssim-motivated rate-distortion optimization for video coding.
\newblock {\em IEEE Transactions on Circuits and Systems for Video Technology},
  22(4):516--529, 2011.

\bibitem{wang2004image}
Zhou Wang, Alan~C Bovik, Hamid~R Sheikh, and Eero~P Simoncelli.
\newblock Image quality assessment: from error visibility to structural
  similarity.
\newblock {\em IEEE transactions on image processing}, 13(4):600--612, 2004.

\bibitem{wang2003foveation}
Zhou Wang, Ligang Lu, and Alan~C Bovik.
\newblock Foveation scalable video coding with automatic fixation selection.
\newblock {\em IEEE Transactions on Image Processing}, 12(2):243--254, 2003.

\bibitem{wang2003multiscale}
Zhou Wang, Eero~P Simoncelli, and Alan~C Bovik.
\newblock Multiscale structural similarity for image quality assessment.
\newblock In {\em The Thrity-Seventh Asilomar Conference on Signals, Systems \&
  Computers, 2003}, volume~2, pages 1398--1402. Ieee, 2003.

\bibitem{wilming2017extensive}
Niklas Wilming, Selim Onat, Jos{\'e}~P Ossand{\'o}n, Alper A{\c{c}}{\i}k, Tim~C
  Kietzmann, Kai Kaspar, Ricardo~R Gameiro, Alexandra Vormberg, and Peter
  K{\"o}nig.
\newblock An extensive dataset of eye movements during viewing of complex
  images.
\newblock {\em Scientific data}, 4(1):1--11, 2017.

\bibitem{winkler1999perceptual}
Stefan Winkler.
\newblock Perceptual distortion metric for digital color video.
\newblock In {\em Human Vision and Electronic Imaging IV}, volume 3644, pages
  175--184. SPIE, 1999.

\bibitem{Yang2020LearningFV}
Ren Yang, Fabian Mentzer, Luc~Van Gool, and Radu Timofte.
\newblock Learning for video compression with hierarchical quality and
  recurrent enhancement.
\newblock {\em 2020 IEEE/CVF Conference on Computer Vision and Pattern
  Recognition (CVPR)}, pages 6627--6636, 2020.

\bibitem{zhang2018deep}
Jing Zhang, Tong Zhang, Yuchao Dai, Mehrtash Harandi, and Richard Hartley.
\newblock {Deep unsupervised saliency detection: A multiple noisy labeling
  perspective}.
\newblock In {\em Proceedings of the IEEE conference on computer vision and
  pattern recognition}, 2018.

\bibitem{zhang2017learning}
Pingping Zhang, Dong Wang, Huchuan Lu, Hongyu Wang, and Baocai Yin.
\newblock Learning uncertain convolutional features for accurate saliency
  detection.
\newblock In {\em Proceedings of the IEEE International Conference on computer
  vision}, 2017.

\bibitem{zhang2021survey}
Yun Zhang, Linwei Zhu, Gangyi Jiang, Sam Kwong, and C-C~Jay Kuo.
\newblock A survey on perceptually optimized video coding.
\newblock {\em arXiv preprint arXiv:2112.12284}, 2021.

\end{thebibliography}
}

\newpage 
\section{Supplementary Material}
\label{sec:sup_material}
\begin{table*}[t]
\centering
\small 
\begin{tabular}{c|ccccc}
\textbf{Input Feature}                                                                                &            &            &            &            &                \\ \hline
Frame                                                                                                   & \checkmark & \checkmark & \checkmark & \checkmark & \checkmark     \\
Saliency                                                                                                & \checkmark & ×          & \checkmark & \checkmark & \checkmark     \\
Object Segmentation                                                                                     & ×          & ×          & \checkmark & \checkmark & \checkmark     \\
Optical Flow                                                                                            & ×          & ×          & ×          & \checkmark & \checkmark     \\
\begin{tabular}[c]{@{}c@{}}Quality Metrics \end{tabular}                                                & ×          & \checkmark & ×          & ×          & \checkmark     \\
\begin{tabular}[c]{@{}c@{}}High Pass Filters\end{tabular}                                               & ×          & ×          & ×          & \checkmark & \checkmark     \\ \hline
\textbf{\begin{tabular}[c]{@{}c@{}}Subjective Study\\  Preference Fraction\end{tabular}}                  & $0.51 \pm 0.04$      & $0.52 \pm 0.04$      & $0.54 \pm 0.04$      & $0.52 \pm 0.04$      & \textbf{$0.67\pm 0.04$}
\end{tabular}
%}
\caption{Ablation Study Results. Subjective Study results shows the fraction of people who preferred predicted PIM-encoded videos over baseline (mean $\pm$ 95\% confidence-interval).}
\label{tab:abelation_study}
\end{table*} 
\setlength\columnsep{19pt}

\subsection{Videos Data}
We used videos from three datasets: 25 videos from CLIC2021 (subset of the UGC dataset \footnote{\url{https://storage.googleapis.com/clic2021_public/txt_files/video_urls.txt}}), 93 videos from BVI-DVC \cite{bvi-dvc}, and 60 videos from Pexels\footnote{\url{https://www.pexels.com}}. Tables  \cref{tab:video_data_clic}, \cref{tab:video_data_bvi_dvc} and \cref{tab:video_data_pexels} show the video name, number of frames, frame rate and bitrate corresponding to an approximate VMAF of 70 for the chosen vidoes in PIMD from CLIC2021, BVI-DVC and Pexels datasets.\\ \\ 

\subsection{Video Bitrates}
\begin{figure}[h]
\centering
\begin{subfigure}[t]{\columnwidth}
\includegraphics[width=\textwidth]{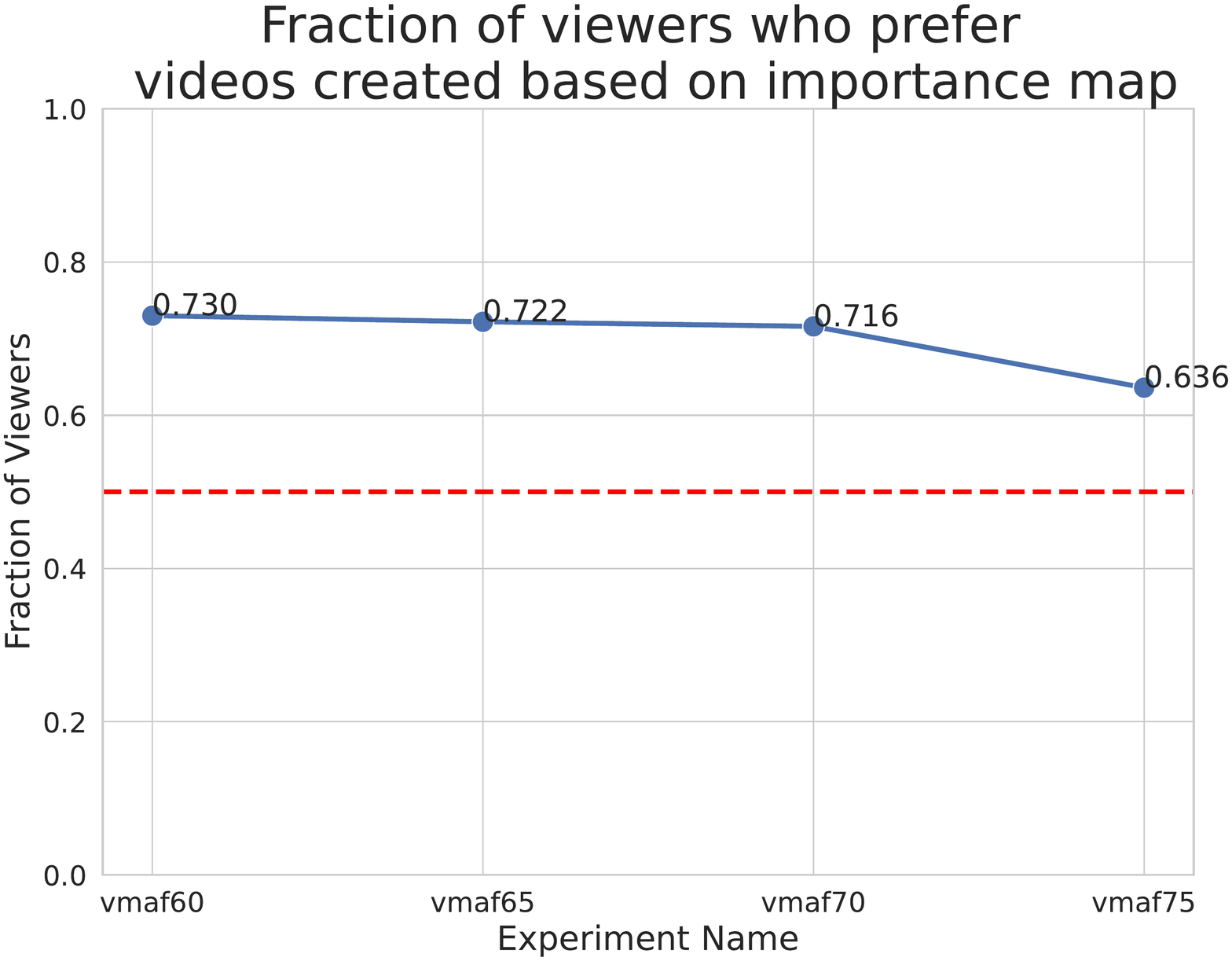}
\end{subfigure}
\caption{We conducted a small pilot study to choose the VMAF for our large scale data collection. The plot shows fraction of viewers who prefer x264 encoded videos based on importance maps compared to x264 encoded baseline videos without importance maps.}
\label{fig:dataset_by_vmaf}
\end{figure}

To select the VMAF for the encoded videos in PIMD, we conducted a two-alternative forced choice (2AFC)
subjective pilot study to understand the trade-off between
bitrate selection and encoding advantages from using the
importance map based encoding. We annotated 25 videos from the CLIC2021 dataset at bitrates that match the VMAF scores of 60, 65, 70 and 75 when compared with the original reference video. This was followed by a two-alternative forced choice (2AFC) subjective study, at each individual VMAF value, in which annotators were asked to indicate their choice between baseline video and video encoded based on the average importance map. The results of this subjective study are shown in  \cref{fig:dataset_by_vmaf}. 

We chose to use bitrates that match a score of VMAF 70, because it is the highest bitrate without a significant drop in performance. Moreover, as mentioned in the paper, these bitrates are low but nevertheless realistic. 

\subsection{Additional Improvements Upon \cite{our-tool}}
In Section 3, in the main paper, we presented several extensions that have been made to the annotation web-tool presented in \cite{our-tool}. We added another feature to the tool before we conducted the large-scale data collection. The video annotation task is an iterative task, with task completion times dependent upon the length of the video, the complexity of the content and the proficiency of the annotators with video compression technologies. Collecting large scale data might become a time consuming task, and we can not expect annotators to annotate in one sitting. Therefore, we have added a modified URL for each pair of video and annotator which allows annotators to directly go to any particular video and continue the task from a video of their choosing.

\subsection{Video Annotation Web-Tool Limitations}
\begin{figure}[H]
\centering
\begin{subfigure}[t]{\columnwidth}
\includegraphics[width=\textwidth]{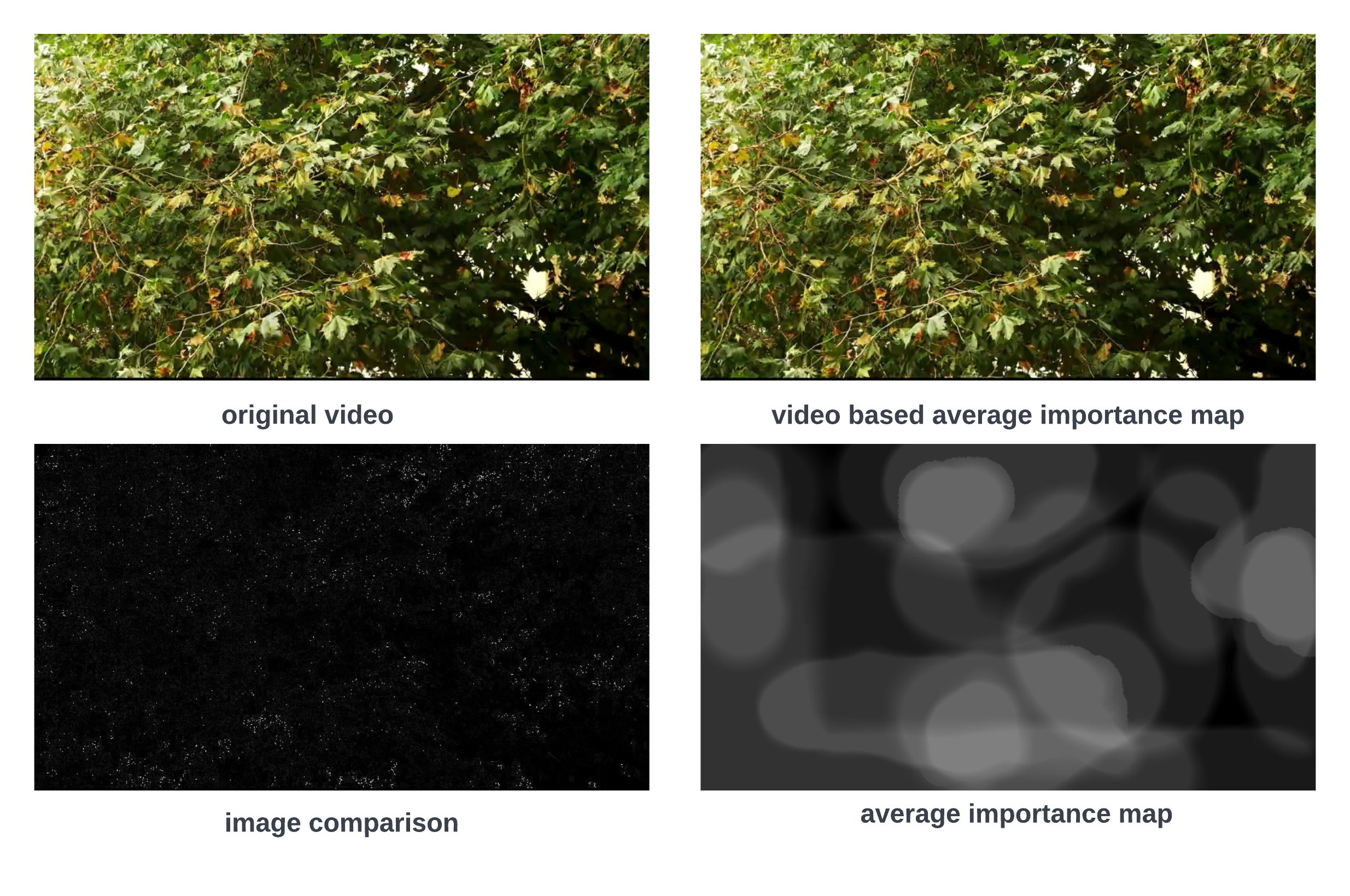}
\end{subfigure}
\caption{An example of a video which is hard to improve using PIMTool.}
\label{fig:tool_limitation}
\end{figure}

Although the video annotation web-tool can be used to improve videos by shifting bits, not all videos seem to carry potential for improvement. \cref{fig:tool_limitation} shows an example of such a video, \emph{ALeaves1BVITexture} from BVI-DVC dataset. The first row shows the first frame from the original video and the first frame from the  video encoded based on the average importance map. Even though annotators tried to annotate this video (average importance map is shown in bottom right of  \cref{fig:tool_limitation}), we see that it leads to no improvement. As can be seen in the image comparison (bottom left in  \cref{fig:tool_limitation}), annotating leads to subtle differences which are not visually prominent. This is expected as looking at the content of this video, there are no obvious regions where a bitrate trade-off would lead to a higher quality.

\subsection{Ablation Studies}

When designing the model, our goal was to provide it with features that would be indicative of the human-annotated importance map. On one hand, some features should highlight areas that an observer would naturally gravitate towards due to the content of the video, but on the other hand others should predict which areas might unintentionally grab one's attention due to compression artifacts. We conducted an ablation study with different subsets of features to understand whether each feature is truly necessary. We trained five different models with same training hyperparameters using different subset of input 
features, and then performed a 2AFC study on the PIMD test-set.  \cref{tab:abelation_study} shows the results for the ablation study. We can see that for all other models except the one containing all the features, we get a chance performance between videos encoded without any importance and PIMM-encoded videos. This highlights the importance of all the chosen features for the model reported in the main text.

% Videos from CLIC2021:
\onecolumn
\begin{longtable}
{|p{7cm}|p{1.5cm}|p{1cm}|p{2.5cm}|}
\hline
\multicolumn{4}{|c|}{CLIC2021 Videos Data  } \\
\hline
video name & \#frames & fps & bitrate (kbps) \\
\hline
\endhead
 \hline\hline
 Animation\_1080P-0cdf & 110 & 25 & 390  \\
 \hline
 CoverSong\_720P-01a1 & 100 & 25 & 80  \\
 \hline
 CoverSong\_720P-449f & 120 & 25 & 70  \\
 \hline
 CoverSong\_1080P-0a86 & 100 & 25 & 105  \\
 \hline
 Gaming\_720P-5ba2 & 110 & 25 & 660  \\
 \hline
 Gaming\_720P-6658 & 120 & 25 & 210  \\
 \hline
 HowTo\_1080P-13aa & 120 & 25 & 210  \\
 \hline
 Lecture\_720P-7dcf & 90 & 25 & 60  \\
 \hline
 LiveMusic\_1080P-14af & 90 & 25 & 240  \\
 \hline
 LiveMusic\_1080P-541f & 80 & 25 & 170  \\
 \hline
 MusicVideo\_1080P-55af & 90 & 25 & 295  \\
 \hline
 Sports\_720P-2c06 & 70 & 25 & 230  \\
 \hline
 Sports\_720P-07d0 & 110 & 25 & 45  \\
 \hline
 Sports\_1080P-3eb0 & 80 & 25 & 260  \\
 \hline
 Sports\_1080P-28a6 & 120 & 25 & 50  \\
 \hline
 Sports\_1080P-43e2 & 75 & 25 & 60  \\
 \hline
 Sports\_1080P-241e & 110 & 25 & 380  \\
 \hline
 Sports\_2160P-2e1d & 120 & 25 & 280  \\
 \hline
 Sports\_2160P-4024 & 90 & 25 & 400  \\
 \hline
 TelevisionClip\_1080P-39e3 & 110 & 25 & 440  \\
 \hline
 TelevisionClip\_1080P-401e & 100 & 25 & 110  \\
 \hline
 TelevisionClip\_1080P-3758 & 100 & 25 & 145  \\
 \hline
 Vlog\_720P-155f & 90 & 25 & 600  \\
 \hline
 Vlog\_720P-329f & 100 & 25 & 690  \\
 \hline
 Vlog\_2160P-09c9 & 90 & 25 & 125  \\ 
 \hline
 \caption{Properties of videos chosen for annotation in PIMD  from \textbf{CLIC2021}. }
\label{tab:video_data_clic}
\end{longtable}

\begin{longtable}{|p{7cm}|p{1.5cm}|p{1cm}|p{2.5cm}|}
\hline
\multicolumn{4}{|c|}{BVI-DVC Videos Data  } \\
\hline
video name & \#frames & fps & bitrate (kbps) \\
\hline
\endhead
 \hline\hline
 AAdvertisingMassagesBangkokVidevo & 64 & 25 & 180  \\
\hline
AAmericanFootballS3Harmonics & 64 & 60 & 415  \\
\hline
ABangkokMarketVidevo & 64 & 25 & 315  \\
\hline
ABasketballGoalScoredS1Videvo & 64 & 25 & 265  \\
\hline
ABasketballGoalScoredS2Videvo & 64 & 25 & 265  \\
\hline
ABoatsChaoPhrayaRiverVidevo & 64 & 23 & 315  \\
\hline
ABoxingPracticeHarmonics & 64 & 60 & 270  \\
\hline
ABuildingRoofS1IRIS & 64 & 24 & 75  \\
\hline
ABuntingHangingAcrossHongKongVidevo & 64 & 25 & 210  \\
\hline
ACatchBVIHFR & 64 & 120 & 590  \\
\hline
ACharactersYonseiUniversity & 64 & 30 & 110  \\
\hline
AChurchInsideMCLJCV & 64 & 30 & 120  \\
\hline
ACityStreetS1IRIS & 64 & 24 & 275  \\
\hline
ACityStreetS6IRIS & 64 & 24 & 275  \\
\hline
ACloseUpBasketballSceneVidevo & 64 & 25 & 465  \\
\hline
AColourfulPaperLanternsVidevo & 64 & 50 & 650  \\
\hline
AColourfulRugsMoroccoVidevo & 64 & 50 & 225  \\
\hline
AConstructionS2YonseiUniversity & 64 & 30 & 125  \\
\hline
ACrosswalkHongKong2S1Videvo & 64 & 25 & 455  \\
\hline
ACrosswalkHongKong2S2Videvo & 64 & 25 & 460  \\
\hline
ACyclistS1BVIHFR & 64 & 120 & 690  \\
\hline
ACyclistVeniceBeachBoardwalkVidevo & 64 & 25 & 300  \\
\hline
ADrivingPOVHarmonics & 64 & 60 & 255  \\
\hline
AEnteringHongKongStallS1Videvo & 64 & 25 & 345  \\
\hline
AFerrisWheelTurningVidevo & 64 & 50 & 695  \\
\hline
AFjordsS1Harmonics & 64 & 60 & 685  \\
\hline
AFlagShootTUMSVT & 64 & 50 & 475  \\
\hline
AFlowerChapelS1IRIS & 64 & 24 & 130  \\
\hline
AFlowerChapelS2IRIS & 64 & 24 & 80  \\
\hline
AFlyingThroughLAStreetVidevo & 64 & 23 & 115  \\
\hline
AGrazTowerIRIS & 64 & 24 & 85  \\
\hline
AHarleyDavidsonIRIS & 64 & 24 & 240  \\
\hline
AHongKongIslandVidevo & 64 & 25 & 395  \\
\hline
AHongKongMarket1Videvo & 64 & 25 & 340  \\
\hline
AHongKongMarket3S1Videvo & 64 & 25 & 420  \\
\hline
AHongKongMarket3S2Videvo & 64 & 25 & 350  \\
\hline
AHongKongMarket4S2Videvo & 64 & 25 & 335  \\
\hline
AHongKongS3Harmonics & 64 & 60 & 560  \\
\hline
AHorseDrawnCarriagesVidevo & 64 & 50 & 445  \\
\hline
AHorseStaringS1Videvo & 64 & 50 & 315  \\
\hline
AHorseStaringS2Videvo & 64 & 50 & 340  \\
\hline
AJockeyHarmonics & 64 & 120 & 650  \\
\hline
AJoggersS2BVIHFR & 64 & 120 & 795  \\
\hline
AKoraDrumsVidevo & 64 & 25 & 420  \\
\hline
ALakeYonseiUniversity & 64 & 30 & 650  \\
\hline
ALungshanTempleS1Videvo & 64 & 50 & 415  \\
\hline
ALungshanTempleS2Videvo & 64 & 50 & 170  \\
\hline
AManMoTempleVidevo & 64 & 25 & 180  \\
\hline
AManStandinginProduceTruckVidevo & 64 & 25 & 295  \\
\hline
AManWalkingThroughBangkokVidevo & 64 & 25 & 85  \\
\hline
AMirabellParkS1IRIS & 64 & 24 & 100  \\
\hline
AMirabellParkS2IRIS & 64 & 24 & 160  \\
\hline
AMobileHarmonics & 64 & 60 & 325  \\
\hline
AMuralPaintingVidevo & 64 & 25 & 275  \\
\hline
AMyanmarS4Harmonics & 64 & 60 & 415  \\
\hline
AMyanmarS6Harmonics & 64 & 60 & 930  \\
\hline
AMyeongDongVidevo & 64 & 25 & 305  \\
\hline
ANewYorkStreetDareful & 64 & 30 & 140  \\
\hline
AOrangeBuntingoverHongKongVidevo & 64 & 25 & 235  \\
\hline
AParkViolinMCLJCV & 64 & 25 & 460  \\
\hline
APedestriansSeoulatDawnVidevo & 64 & 25 & 265  \\
\hline
APersonRunningOutsideVidevo & 64 & 50 & 220  \\
\hline
ARollerCoaster2Netflix & 64 & 60 & 695  \\
\hline
ARunnersSJTU & 64 & 60 & 805  \\
\hline
ASeasideWalkIRIS & 64 & 24 & 255  \\
\hline
ASeekingMCLV & 64 & 25 & 695  \\
\hline
AShoppingCentreVidevo & 64 & 25 & 260  \\
\hline
ASignboardBoatLIVENetFlix & 64 & 30 & 175  \\
\hline
AStreetDancerS1IRIS & 64 & 24 & 185  \\
\hline
AStreetDancerS2IRIS & 64 & 24 & 235  \\
\hline
AStreetDancerS3IRIS & 64 & 24 & 230  \\
\hline
AStreetDancerS4IRIS & 64 & 24 & 230  \\
\hline
AStreetDancerS5IRIS & 64 & 24 & 210  \\
\hline
AStreetsOfIndiaS2Harmonics & 64 & 60 & 355  \\
\hline
ATaiChiHongKongS1Videvo & 64 & 25 & 215  \\
\hline
ATaiChiHongKongS2Videvo & 64 & 25 & 225  \\
\hline
ATaipeiCityRooftops8Videvo & 64 & 25 & 150  \\
\hline
ATaksinBridgeVidevo & 64 & 23 & 155  \\
\hline
ATennisMCLV & 64 & 24 & 285  \\
\hline
AToddlerFountain2Netflix & 64 & 60 & 1990  \\
\hline
ATouristsSatOutsideVidevo & 64 & 25 & 230  \\
\hline
AToyCalendarHarmonics & 64 & 60 & 535  \\
\hline
ATrackingDownHongKongSideVidevo & 64 & 25 & 155  \\
\hline
ATrackingPastRestaurantVidevo & 64 & 25 & 305  \\
\hline
ATrafficFlowSJTU & 64 & 60 & 350  \\
\hline
ATunnelFlagS1Harmonics & 64 & 60 & 820  \\
\hline
AUnloadingVegetablesVidevo & 64 & 25 & 300  \\
\hline
AVegetableMarketS1LIVENetFlix & 64 & 30 & 270  \\
\hline
AVegetableMarketS2LIVENetFlix & 64 & 30 & 295  \\
\hline
AVegetableMarketS4LIVENetFlix & 64 & 30 & 330  \\
\hline
AVeniceS1Harmonics & 64 & 60 & 95  \\
\hline
AWalkingDownKhaoStreetVidevo & 64 & 25 & 215  \\
\hline
AWalkingDownNorthRodeoVidevo & 64 & 25 & 300  \\
\hline
\caption{Properties of videos chosen for annotation in PIMD  from \textbf{BVI-DVC}. }
\label{tab:video_data_bvi_dvc}
\end{longtable}

\begin{longtable}{|p{7cm}|p{1.5cm}|p{1cm}|p{2.5cm}|}
\hline
\multicolumn{4}{|c|}{Pexels Videos Data  } \\
\hline
video name & \#frames & fps & bitrate (kbps) \\
\hline
\endhead
 \hline\hline
853934 & 100 & 25 & 30  \\
\hline
854136 & 100 & 25 & 50  \\
\hline
855418 & 100 & 25 & 60  \\
\hline
855563 & 100 & 25 & 115  \\
\hline
855574 & 100 & 25 & 75  \\
\hline
855770 & 100 & 25 & 180  \\
\hline
856166 & 100 & 25 & 150  \\
\hline
979689 & 100 & 25 & 160  \\
\hline
992592 & 100 & 25 & 190  \\
\hline
992695 & 100 & 25 & 170  \\
\hline
1085999 & 100 & 25 & 25  \\
\hline
1154850 & 100 & 25 & 140  \\
\hline
1249406 & 100 & 25 & 230  \\
\hline
1580507 & 100 & 25 & 215  \\
\hline
1580509 & 100 & 25 & 130  \\
\hline
1596891 & 100 & 25 & 60  \\
\hline
1943413 & 100 & 25 & 155  \\
\hline
2086113 & 100 & 25 & 60  \\
\hline
2099568 & 100 & 25 & 325  \\
\hline
2675511 & 100 & 25 & 120  \\
\hline
2828735 & 100 & 25 & 20  \\
\hline
3015527 & 100 & 25 & 370  \\
\hline
3018533 & 100 & 25 & 225  \\
\hline
3018669 & 100 & 25 & 55  \\
\hline
3064217 & 100 & 25 & 175  \\
\hline
3115237 & 100 & 25 & 40  \\
\hline
3115442 & 100 & 25 & 215  \\
\hline
3135808 & 100 & 25 & 165  \\
\hline
3201794 & 100 & 25 & 30  \\
\hline
3205917 & 100 & 25 & 180  \\
\hline
3206478 & 100 & 25 & 70  \\
\hline
3326928 & 100 & 25 & 365  \\
\hline
3444430 & 100 & 25 & 105  \\
\hline
3444432 & 100 & 25 & 70  \\
\hline
3770034 & 100 & 25 & 145  \\
\hline
3770054 & 100 & 25 & 155  \\
\hline
3773486 & 100 & 25 & 90  \\
\hline
3779922 & 100 & 25 & 80  \\
\hline
3873059 & 96 & 25 & 175  \\
\hline
3987730 & 100 & 25 & 195  \\
\hline
4089575 & 100 & 25 & 170  \\
\hline
4166533 & 100 & 25 & 80  \\
\hline
4174346 & 100 & 25 & 170  \\
\hline
4221451 & 100 & 25 & 155  \\
\hline
4438080 & 100 & 25 & 275  \\
\hline
4443257 & 100 & 25 & 90  \\
\hline
4446327 & 100 & 25 & 50  \\
\hline
4480576 & 100 & 25 & 85  \\
\hline
4782855 & 100 & 25 & 65  \\
\hline
5236593 & 100 & 25 & 190  \\
\hline
5583559 & 100 & 25 & 65  \\
\hline
5607553 & 100 & 25 & 215  \\
\hline
5645037 & 100 & 25 & 125  \\
\hline
6563849 & 100 & 25 & 160  \\
\hline
7334679 & 100 & 25 & 255  \\
\hline
7502873 & 100 & 25 & 420  \\
\hline
7613411 & 100 & 25 & 85  \\
\hline
7709982 & 100 & 25 & 105  \\
\hline
7782528 & 100 & 25 & 265  \\
\hline
7975463 & 100 & 25 & 25  \\
\hline 
\caption{Properties of videos chosen for annotation in PIMD  from \textbf{Pexels}.}
\label{tab:video_data_pexels}
\end{longtable}

\end{document}